\documentstyle[pra,aps]{revtex}
\tightenlines
\begin{document}

\twocolumn[\hsize\textwidth\columnwidth\hsize
\csname@twocolumnfalse%
\endcsname

\def\beq{\begin{equation}}
\def\eeq{\end{equation}}
\def\bea{\begin{eqnarray}}
\def\eea{\end{eqnarray}}
\def\bv{{\bf v}}
\def\br{{\bf r}}

\draft

\title{Sound propagation in a cylindrical Bose-condensed gas}

\author{E. Zaremba}
\address{Department of Physics, Queen's University \\ Kingston,
Ontario, Canada K7L 3N6}

\date{\today}

\maketitle

\begin{abstract}
We study the normal modes of a cylindrical Bose condensate 
at $T = 0$ using
the linearized time-dependent Gross-Pitaevskii equation in the
Thomas-Fermi limit. These modes are relevant to the recent
observation of pulse propagation in long, cigar-shaped traps. We
find that pulses generated in a cylindrical condensate propagate
with little spread at a speed $c = \sqrt{g\bar n /m}$, where
$\bar n$ is the average density of the condensate over its
cross-sectional area.
\end{abstract}

]

In a recent paper, Andrews {\it et al.}\cite{andrews97}
presented results for the propagation of sound through the Bose
condensate in an axially-symmetric, cigar-shaped trap. These
observations were made possible by the large aspect ratio of the
anisotropic trap and the localized generation of pulses by a
laser beam focused at the mid-point along the length of the
trap. By imaging the propagation of the pulse, they were able to
determine the sound speed as a function of the condensate
density. The measured sound speed is approximately equal to the
Bogoliubov phonon velocity as determined by the maximum density
in the cloud. The authors\cite{andrews97} sketch a qualitative 
theoretical argument in support of this observation.

The purpose of the present paper is to provide a more detailed
analysis of the dependence of the sound speed on the
inhomogeneous distribution of atoms in the trap. In the
propagation of a pulse for short
times, the ends of the trapped condensate do not come into play
and we can therefore simulate the experimental situation by
considering an idealized cylindrical trap which is uniform in the
$z$-direction and which has a harmonic confining potential of the form
\begin{equation}
V(\rho) = {1\over 2} m \omega_0^2 \rho^2
\end{equation}
in the radial direction. Here, $\omega_0$ is the trap
frequency. Treating the condensate in the Thomas-Fermi (TF)
approximation\cite{baym96}, the equilibrium density is given by
\begin{equation}
n_0(\rho) = {m\omega_0^2 \over 2g}(R^2 - \rho^2)\,,
\label{density}
\end{equation}
where $g = 4\pi a\hbar^2/m$ is the interaction parameter and $R$
is the radius of the cylindrical condensate related to the
chemical potential, $\mu=V(\rho) + gn_0(\rho)$, 
by $\mu = {1\over 2}m\omega_0^2 R^2$.

The dynamics of the condensate will be based on the linearized
time-dependent Gross-Pitaevskii equation\cite{pitaevskii61} in
the TF limit\cite{stringari96}. This equation can be recast
as a pair of quantum hydrodynamic equations
\begin{eqnarray}
{\partial \delta n \over \partial t} &=& - \nabla\cdot (n_0 {\bf v}) \\
m{\partial {\bf v} \over \partial t} &=& - g \nabla \delta n 
-\nabla \delta U\,,
\label{velocity} 
\end{eqnarray}
where $\delta U({\bf r},t)$ represents an externally imposed 
potential. Eliminating the velocity from these equations, we obtain 
an equation for the density fluctuation $\delta n({\bf r},t)$ first
derived (for $\delta U = 0$) by Stringari\cite{stringari96},
\begin{equation}
{\partial^2 \delta n \over \partial t^2} = {g \over m} 
\nabla \cdot (n_0 \nabla \delta n) + {1\over m} \nabla \cdot
(n_0\nabla \delta U)\,.
\label{dyneqn}
\end{equation}

To study the normal modes of the condensate, we set $\delta U$
to zero and look for solutions which correspond to
a propagating wave of the form
\begin{equation}
\delta n({\bf r},t) = \delta n(\rho) e^{i(kz -\omega t)}\,.
\label{deltan}
\end{equation}
We assume that the density fluctuation in the transverse
directions depends only on the radial variable $\rho$ and not on the
azimuthal angle $\phi$. It is clear that more general solutions
having the angular dependence $e^{im\phi}$ are possible, but
these will not be considered in this paper.

We now substitute this form of the solution into (\ref{dyneqn})
and, making use of the equilibrium density in (\ref{density}),
obtain the equation ($\delta U=0$)
\begin{eqnarray} 
\omega^2 \delta n = {1\over 2}\omega_0^2 \Bigg \{ &k^2&(R^2 -
\rho^2)\delta n - {1\over \rho}(R^2 -3\rho^2){\partial \delta n
\over \partial \rho} \nonumber \\
&-& (R^2 -\rho^2) {\partial^2 \delta n \over
\partial \rho^2} \Bigg \}\,,
\end{eqnarray}
where $\delta n$ now represents the spatially-dependent
amplitude defined in (\ref{deltan}). The allowed solutions of
this equation define the 
dispersion relations, $\omega (k)$, of the various
modes of the condensate in the cylindrical trap.

It is now convenient to introduce the dimensionless parameters
\begin{displaymath}
\bar \omega \equiv {\omega \over \omega_0},\qquad
\bar k \equiv kR 
\end{displaymath}
and to define the new independent variable
\begin{equation}
x \equiv 2{\rho^2 \over R^2} -1,\qquad -1 \le x \le 1\,.
\end{equation}
When expressed in terms of this variable, the density
fluctuation, now denoted by $y(x)$, is found to satisfy
\begin{equation}
{d \over dx} \left [ (1-x^2) {dy \over dx} \right ] + {1\over 2}
\bar \omega^2 y - {1\over 8} \bar k^2 (1-x) y = 0\,,
\label{yeqn}
\end{equation}
which is in the standard Sturm-Liouville form. For $\bar k = 0$,
this equation in fact reduces to the Legendre differential
equation 
\begin{equation}
{d \over dx} \left [ (1-x^2) {dy \over dx} \right ] + \lambda y =
0\,,
\end{equation}
whose solutions are the Legendre polynomials $P_l(x)$ with
eigenvalues $ \lambda = l(l+1)$, $l = 0,1,\ldots$ Thus the modes
of the condensate can be labeled by the index $l$ and 
have a limiting frequency at $k =0$ given by
\begin{equation}
\omega_l^2(k=0) = 2l(l+1)\omega_0^2\,.
\label{sequence}
\end{equation}
The first two frequencies in the sequence, 0 and $2\omega_0$, are
the two $m=0$ mode frequencies found by
Stringari\cite{stringari96} for a cigar-shaped trap in the limit
that the axial frequency, $\omega_z$, tends to zero. More
generally, the whole sequence in (\ref{sequence}) was recently
obtained analytically in this same limit by Fliesser {\it et
al.}\cite{fliesser97}.

These various solutions have distinct radial eigenfunctions. For
example, for $l =0$, $y(x)=$ a constant, and the associated
density fluctuation has a radially independent amplitude. This
mode corresponds to a local adiabatic expansion of the
condensate brought about by a local change in the chemical 
potential, $\delta \mu$. Such a change gives rise to a
variation in the equilibrium density $n_0(\rho;R)$ which is
spatially independent. In the limit of
long wavelengths, it costs no energy to move atoms from one
region of the trap to another and the mode therefore has zero
frequency. The other $l \ne 0$ modes have a finite frequency
since they involve radial motion of the condensate. (Note that
according to (\ref{velocity}), the direction of the local 
velocity is normal to constant density surfaces.)
For example,
for $l=1$, the density fluctuation is of the form $\delta
n(\rho) \propto \left ( 1 - 2\rho^2/R^2 \right )$ which
has a node at $\rho = R/\sqrt{2}$. Unlike the $l=0$ mode, all of
these modes have a number conserving density fluctuation in the
sense that $\int_0^R d\rho \, \rho\, \delta n(\rho) = 0$. The $l=1$
mode at $k=0$ is a radial breathing mode.

For $k \ne 0$, we expand the density fluctuation as
\begin{equation}
y(x) = \sum_l a_l P_l(x)\,,
\label{expand}
\end{equation}
where the normalization of the Legendre functions is chosen
such that
\begin{equation}
\int_{-1}^1 dx P_l(x) P_{l'}(x) = \delta_{ll'}\,.
\end{equation}
Substituting expansion (\ref{expand}) into (\ref{yeqn}), we obtain 
the system of linear equations
\begin{equation}
\left ( {1\over 2} \bar \omega^2 - l(l+1) -{1\over 8}\bar k^2
\right ) a_l + {1\over 8}\bar k^2 \sum_{l'} M_{ll'} a_{l'} =
0\,,
\label{matrix}
\end{equation}
where the symmetric matrix $M_{ll'}$ is given by
\begin{equation}
M_{ll'} = \int_{-1}^1 dx P_l(x)  x P_{l'}(x)\,.
%&=& \cases{{ \textstyle{l+1}\over 
%\textstyle{\sqrt{(2l+1)(2l+3)}}}, &if
%$l'=l+1$\cr{\textstyle{l}\over \textstyle{\sqrt{(2l-1)(2l+1)}}}, 
%&if $l'=l-1\,.$\cr}
\end{equation}
Since the matrix
$M_{ll'}$ does not have diagonal matrix elements, it is clear
that its effect on the mode eigenvalues first occurs in second
order, so that to lowest order in $k^2$,
\begin{equation}
\omega_l^2(k) = 2l(l+1)\omega_0^2 + {1\over 4}(kR\omega_0)^2 +
O(k^4)\,.
\end{equation}
In particular, the $l=0$ mode is phonon-like with a velocity
\begin{equation}
c = {1\over 2} R \omega_0 = \sqrt{{gn_0(0) \over 2m}}\,,
\label{c}
\end{equation}
where $n_0(0)$ is the (maximum) condensate density on the axis of the
trap. If we compare this velocity with the Bogoliubov velocity
$\sqrt{gn/m}$ in a homogeneous gas of density $n$, we see that
the effective density in the cylindrical trap is half the
maximum value, which is just the density, $\bar n$, obtained by
averaging (\ref{density}) over the condensate cross-section.

It is interesting to note that the matrix problem in
(\ref{matrix}) is equivalent to the problem of a quantum rigid
rotor in a gravitational field, having the Hamiltonian
\begin{equation}
H = {L^2 \over 2I} + MgR(1-\cos \theta)\,.
\end{equation}
Here, $L$ is the angular momentum operator, $I = M R^2$ is
the moment of inertia of the rotor
and the gravitational potential energy of the rotor mass $M$
is referenced with respect to its lowest
position (the polar angle $\theta$ is defined with respect
to this position). The angular eigenfunctions of this
Hamiltonian can be expanded in spherical harmonics
$Y_{lm}(\theta,\phi)$ and in the $m=0$ subspace, the problem is
then identical to (\ref{matrix}). Thus the parameter $\bar k$
plays the role of a `gravitational field' which perturbs the
mode energies from their `free rotor' values. 

The solution of (\ref{matrix}) is easily obtained for arbitrary
$\bar k$ by numerical diagonalization. In Fig. 1 we show some of
the lowest mode frequencies as a function of wavevector. The
lowest mode is a sound-like mode and exhibits a negative
dispersion. It can be seen that the group velocity deviates
appreciably from its long wavelength limit once the wavelength
is comparable to the diameter of the condensate. The
mode-coupling induced by the $\bar k^2$ perturbation in
(\ref{matrix}) of course becomes more significant with
increasing $\bar k$ and has the effect of lowering the sound
speed. This coupling is associated with the interplay of the
wave-like modulation of the density along the axis and the strong
inhomogeneity of the equilibrium density in the radial
direction.

We now return to the experimental situation\cite{andrews97}
which involves the
propagation of sound pulses rather than continuous waves. These
pulses are generated by switching on a laser beam which repels
atoms from its point of application at the center of the
cylindrical trap. We shall assume this perturbation to be weak
and consider the linear response of the condensate. The
perturbation is taken to be
\begin{equation}
\delta U({\bf r},t) = U_0 e^{-z^2/\sigma^2} \theta(t)\,,
\end{equation}
that is, a gaussian potential with no radial dependence, which is
switched on at $t =0$. The equation defining the density
fluctuation is given by (\ref{dyneqn}) and can be solved by
introducing a Fourier representation of
the density and external potential:
\begin{eqnarray}
\delta n(\rho,z,t) &=& \int_{-\infty}^\infty {dk \over 2\pi}
e^{ikz} \delta n(\rho,k,t) \nonumber \\
\delta U(z,t) &=& \int_{-\infty}^\infty {dk \over 2\pi}
e^{ikz} \delta U(k,t)\,,
\end{eqnarray}
with
\begin{equation}
\delta U(k,t) = \sqrt{\pi} \sigma U_0 e^{-\sigma^2k^2/4}
\theta(t) \equiv \delta U(k) \theta(t)\,.
\end{equation}
Taking the Fourier transform of (\ref{dyneqn}) and using the
variable $x$ introduced earlier, we obtain the equation
%, we obtain the
%following equation for the Fourier amplitude $\delta
%n(\rho,k,t)$:
%\begin{eqnarray} 
%{\partial^2 \delta n \over \partial t^2} &=& 
%{1\over 2}\omega_0^2 \Bigg \{ 
%{1\over \rho}(R^2 -3\rho^2){\partial \delta n
%\over \partial \rho} + (R^2 -\rho^2) {\partial^2 \delta n \over
%\partial \rho^2} \nonumber \\ 
%&-& k^2(R^2 - \rho^2)\delta n \Bigg \}
%- {\omega_0^2 k^2 \over 2g} (R^2 - \rho^2) \delta U \,.
%\end{eqnarray}
%Using the variable $x$ introduced earlier, this equation can be
%written as
\begin{equation}
{\partial^2 y \over \partial t^2} = -\hat L y - {(\omega_0kR)^2
\over 4g} (1-x) \delta U
\label{yeqn2}
\end{equation}
where we have defined the differential operator
\begin{equation}
\hat L \equiv -2\omega_0^2 \left [ {d \over dx} (1-x^2)
{d \over dx} - {1\over 8} \bar k^2(1-x) \right ]\,.
\end{equation}
This is just the differential operator defining the mode
eigenfunctions in (\ref{yeqn}). These eigenfunctions satisfy
\begin{equation}
\hat L \phi_m(x) = \omega_m^2(k) \phi_m(x)
\end{equation}
and form an orthonormal set,
\begin{equation}
\int_{-1}^1 dx \phi_m(x) \phi_n(x) = \delta_{mn}\,.
\end{equation}
They can therefore be used to expand $y(x,k,t)$ as
\begin{equation}
y(x,k,t) = \sum_m b_m(k,t) \phi_m(x)\,.
\end{equation}
Substituting this expansion into (\ref{yeqn2}) and taking the inner
product of the resulting equation with $\phi_n(x)$, we obtain
the equation for a driven harmonic oscillator
\begin{equation}
\ddot b_n + \omega_n^2 b_n = f_n(t)\,,
\label{begin{equation}n}
\end{equation}
where the forcing term on the right hand side is defined as
\begin{eqnarray}
f_n(t) &=& - {(\omega_0 k R)^2 \over 4g} \delta U(k,t) \int_{-1}^1
dx \phi_n(x) (1-x) \nonumber \\ &\equiv& f_n(k) \theta(t)\,.
\end{eqnarray}
The solution of (\ref{begin{equation}n}) for $t \ge 0$ with the boundary
conditions $b_n(0) = 0$ and $\dot b_n(0) = 0$ is
\begin{equation}
b_n(k,t) = {f_n(k) \over \omega_n^2(k)} [1 - \cos \omega_n(k) t]\,.
\label{bn}
\end{equation}
We have here explicitly displayed the dependence of the various
quantities on the wavevector $k$ in the Fourier expansion. This
essentially completes the solution for the density fluctuation
$\delta n(\rho,z,t)$.

To analyze the time evolution of the density pulse, it is
convenient to consider the average of $\delta n(\rho,z,t)$ over
the cross-sectional area, $A$, of the condensate. We therefore define
\begin{eqnarray}
\overline{\delta n}(z,t) &=& {1\over A}\int_A dA\, \delta n(\rho,z,t)
\nonumber \\
&=& \int_{-\infty}^\infty {dk \over 2\pi} e^{ikz} \sum_n
b_n(k,t) {1\over 2} \int_{-1}^1 dx \phi_n(x)\,.
\end{eqnarray}
According to (\ref{expand}), the
eigenfunctions $\phi_n(x)$ are themselves given by the
expansion
\begin{equation} \phi_n(x) = \sum_l a_l^{(n)}(k) P_l(x)\,,
\end{equation}
where the coefficients $a_l^{(n)}(k)$ defining the $n$-th 
eigenvector of (\ref{matrix}) have the limiting value
$a_l^{(n)}(k=0) = \delta_{ln}$. (In other words, the eigenfunction
$\phi_n(x)$ reduces to $P_n(x)$ in the $k \to 0$ limit.)
The orthonormality of the Legendre functions thus yields
\begin{equation} 
\overline{\delta n}(z,t) = \int_{-\infty}^\infty {dk \over 2\pi}
e^{ikz} {1\over \sqrt{2}} \sum_n a_0^{(n)}(k) b_n(k,t)\,.
\label{avgden}
\end{equation}
Similarly, we find
\begin{equation}
f_n(k) = - {(\omega_0 k R)^2 \over 2\sqrt{2} g} \left ( a_0^{(n)}(k)
- {1\over \sqrt{3}} a_1^{(n)}(k) \right ) \delta U(k)\,.
\label{fn}
\end{equation}
Thus, for small $k$, only the $n=0$ and $n=1$ modes are excited
appreciably.

Substituting (\ref{bn}) together with (\ref{fn}) into
(\ref{avgden}), the averaged density fluctuation can be
evaluated numerically. As an example of the calculation, we
present in Fig. 2 results for the propagation of a pulse. The width 
of the gaussian perturbation in these calculations was chosen to be
$\sigma = 1.5R$, which corresponds approximately to the
experimental situation. Once the pulse is launched it 
is seen to propagate
at the speed $c$ given by (\ref{c}). The shape of the pulse
will change in time as a result of the $k$-dispersion of the
normal modes, but the effect is weak since the pulse is mainly
made up of the long-wavelength $n= 0$ sound modes.
The particles carried off by the pulse
leave behind a static depression at the origin which 
corresponds to the new equilibrium shape of the condensate
in the presence of the applied localized perturbation.

These theoretical results for the propagation of a pulse are in 
reasonable agreement with the experimental 
observations\cite{andrews97}. One difference is the stronger
wave packet dispersion observed experimentally which may be due to the
actual variation of the condensate density along the trap axis.
However, a more serious discrepancy concerns the speed of
propagation. In the experiment it was concluded that the sound
speed was determined by the maximum condensate density according
to $c' \simeq \sqrt{gn_0(0)/m}$, as opposed to the value we find
in (\ref{c}). However it is clear from the published figure
that the measured sound speed
actually lies below $c'$ at the higher trap densities. At these
densities, the theoretical value $c$ would lie below the
measured values by about the same amount that $c'$ lies above.  
The difference between our theoretical 
result and experiment could be explained
by errors in the determination of the trap density on which the 
theoretical value is based, but the claimed
experimental precision (W. Ketterle, private communication)
would seem to eliminate this possibility. 

On the theoretical side, one can question
the validity of the TF approximation. 
For the given trap parameters\cite{andrews97} and a typical
density of $n_0(0) \sim 3\times 10^{20}$ m$^{-3}$, the radius of
the condensate $R$, in units of the oscillator length $l =
\sqrt{\hbar/m\omega_0}$, is approximately 5. This implies that
the TF approximation is providing a good estimate of the ground
state density over most of the occupied volume. The validity of
the TF approximation is further supported by the fact that
the frequencies of the
low-lying collective excitations in these traps\cite{mewes96} 
are close to the TF 
values\cite{stringari96}. This observation is consistent with
calculations based on the Bogoliubov 
approximation\cite{singh96,edwards96,perez96}
which show a reasonably rapid convergence to the TF limit with
increasing $N$, the number of trapped atoms.
However, the situation with regard to pulse propagation is
somewhat different in that shorter wavelength excitations are
being probed.
Within the hydrodynamic formulation\cite{stringari96} of 
the Bogoliubov approximation,
an additional term appears on the right hand side of
(\ref{velocity}) which represents the
fluctuation in the `quantum pressure' associated with the
kinetic energy. This term becomes increasingly important as the
wavelength of the density fluctuation is reduced. Unfortunately,
it cannot be estimated using the TF density since the failure
of the TF approximation at the edge of the 
condensate\cite{baym96,dalfovo96} leads to a divergent result. 
Thus, explicit numerical solutions of the Bogoliubov equations
are required in order to ascertain the accuracy of the TF 
approximation for the conditions under which the pulse 
propagation experiments were performed. On the other hand, at 
long wavelengths, we
expect the TF prediction for the sound speed, as given by (\ref{c}),
to be valid.

This work was supported by a grant from 
the Natural Sciences and Engineering Research Council of Canada.
I would like to thank Dr. A. Griffin for a critical reading of
the manuscript and Dr. W. Ketterle for useful discussions.
I would also like to acknowledge the hospitality of Il Ciocco
where this work was begun.

%%%%%%

\centerline{\bf FIGURE CAPTIONS}
\begin{itemize}
\item[Fig.1:]
The dispersion of the lowest modes in a cylindrical condensate
as a function of the
wavevector $k$. $\omega_0$ is the radial trap frequency and $R$
is the TF condensate radius.

\item[Fig.2:]
The propagation of a pulse generated by a gaussian perturbation
applied at the origin at $t =0$. The
lowest curve is at a time $t = 4/\omega_0$, and the interval
between each successive curve is $\Delta t = 4/\omega_0$.

\end{itemize}

\begin{references}
\bibitem{andrews97} M.R. Andrews, D.M. Kurn, H.-J. Miesner, D.S.
Durfee, C.G. Townsend, S. Inouye and W. Ketterle, \prl {\bf 79},
553 (1997).

\bibitem{baym96} G. Baym and C.J. Pethick, \prl {\bf 76}, 6
(1996).

\bibitem{pitaevskii61} L.P. Pitaevskii, Zh. Eksp. Teor. Fiz.
{\bf 40}, 646 (1961) [Sov. Phys. JETP {\bf 13}, 451 (1961)];
E.P. Gross, Nuovo Cimento {\bf 20}, 454 (1961); J. Math. Phys.
{\bf 4}, 195 (1963).

\bibitem{stringari96} S. Stringari, \prl {\bf 77}, 2360 (1996).

\bibitem{fliesser97} M. Fliesser, A. Csord\'as, P. Sz\'epfalusy
and R. Graham, preprint, cond-mat/9706002.

\bibitem{mewes96} M.-O. Mewes, M.R. Andrews, N.J. van Druten,
D.M. Kurn, D.S. Durfee, C.G. Townsend and W. Ketterle, \prl {\bf
77}, 988 (1996).

\bibitem{singh96} K. G. Singh and D. S. Rokhsar, Phys. Rev. Lett.
{\bf 77}, 1667 (1996).

\bibitem{edwards96} M. Edwards, P.A. Ruprecht, K. Burnett, R.J.
Dodd and C.W. Clark, Phys. Rev. Lett. {\bf 77}, 1671 (1996).

\bibitem{perez96} V.M. P\'erez-Garc\'{\i}a, H. Michinel, J.I.
Cirac, M. Lewenstein and P. Zoller, \prl {\bf 77}, 5320 (1996).

\bibitem{dalfovo96} F. Dalfovo, L.P. Pitaevskii and S.
Stringari, Phys. Rev. A {\bf 54}, 4213 (1996).

\end{references}
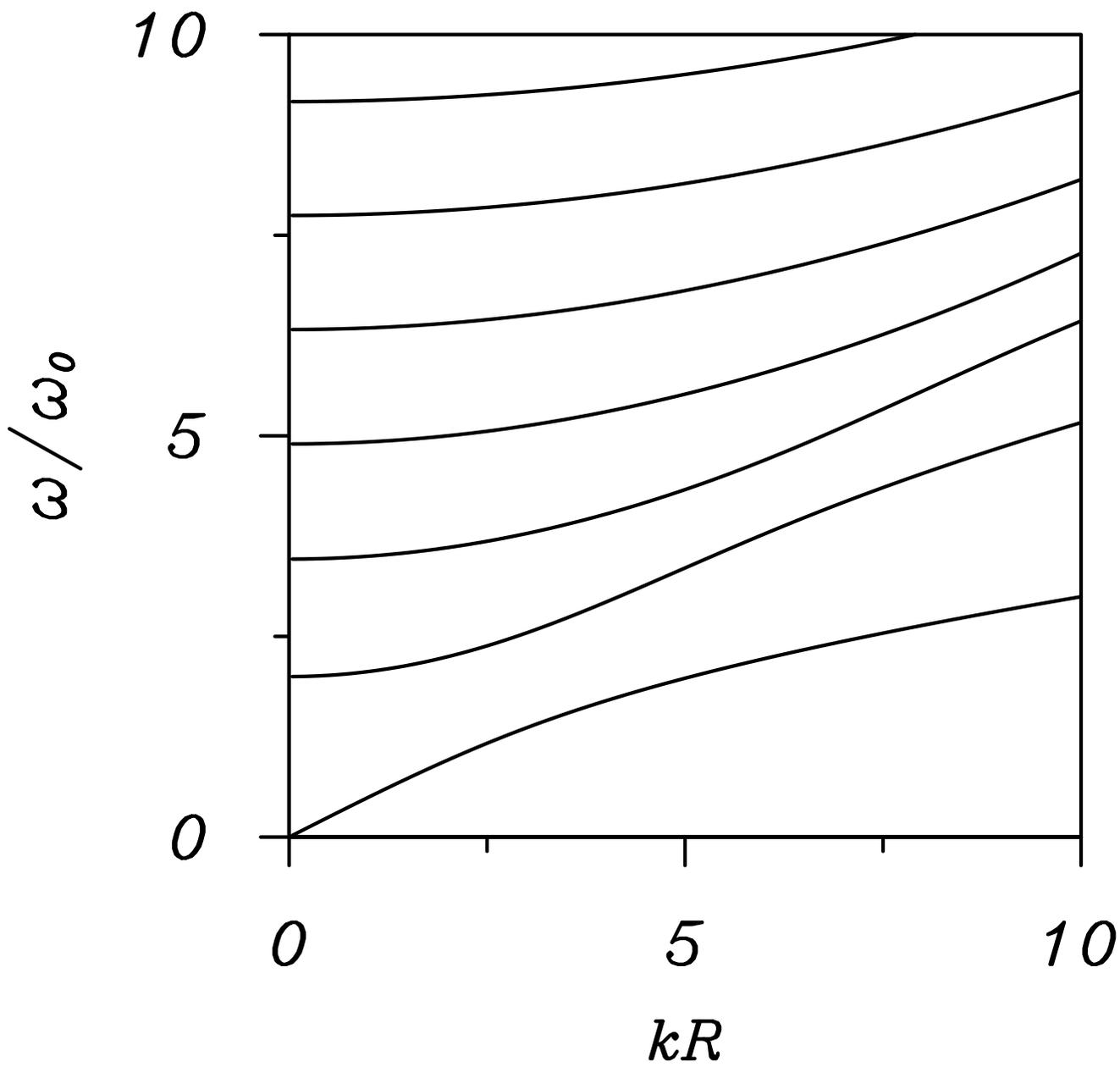
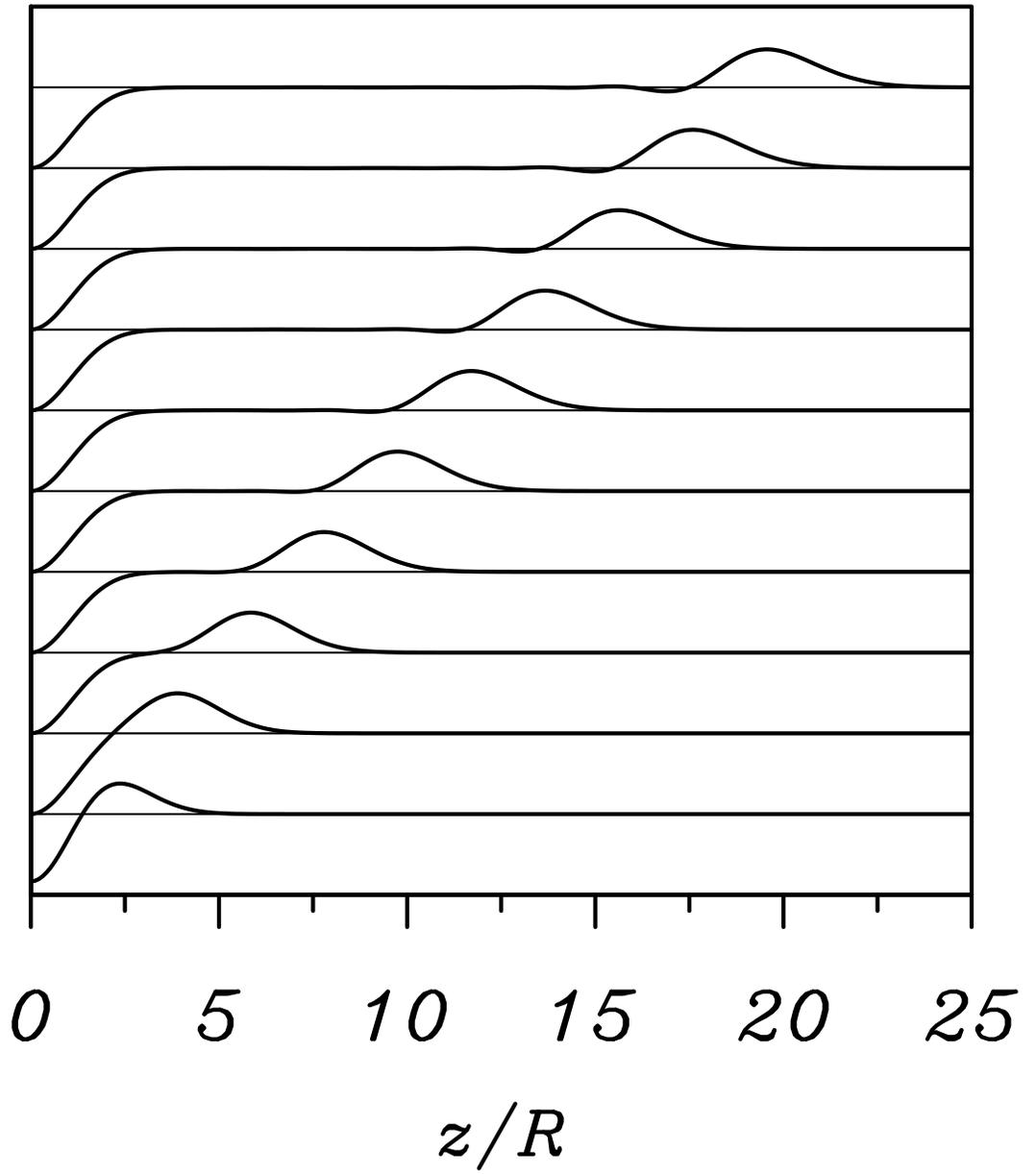
\end{document}